\documentclass[letterpaper,12pt]{article}
\usepackage[margin=1in,letterpaper]{geometry}
\usepackage{amssymb}
\usepackage{amsmath}
\usepackage[hyphens]{url}
\usepackage{listings}
\usepackage{setspace}
\usepackage{subcaption}
\usepackage{graphicx}
\usepackage{float}
\usepackage{natbib}
\usepackage{booktabs}
\usepackage{url}
\usepackage{chngcntr}
\usepackage{breakcites}
\usepackage{multirow}
\usepackage{xcolor}
\usepackage{caption}
\usepackage{mathtools}
\usepackage{relsize}
\usepackage{mathptmx}

\title{Automated classification for open-ended questions with BERT}
\date{}

\author{Hyukjun Gweon\thanks{Department of Statistical and Actuarial Sciences, Western University, London, Ontario, N6A 5B7, Canada}
   \and
   Matthias Schonlau\thanks{Department of Statistics and Actuarial Science, University of Waterloo, Waterloo, Ontario, N2L 3G1, Canada}
}

\raggedright
\begin{document}

\maketitle

\begin{abstract}
Manual coding of text data from open-ended questions into different categories is time consuming and expensive.  Automated coding uses statistical/machine learning  to train on  a small subset of manually coded text answers. Recently, pre-training a general language model on vast amounts of unrelated data and then adapting the model to the specific application has proven effective in natural language processing. Using two data sets, we empirically investigate whether BERT, the currently dominant pre-trained language model, is more effective at  automated coding of answers to open-ended questions than other non-pre-trained statistical learning approaches. We found fine-tuning the pre-trained BERT parameters is essential as otherwise BERT's is not competitive. Second, we found fine-tuned BERT barely beats the non-pre-trained statistical learning approaches in terms of classification accuracy when trained on 100 manually coded observations. However, BERT's relative advantage increases rapidly when more manually coded observations (e.g. 200-400) are available for training. We conclude that for automatically coding answers to open-ended questions BERT is preferable to non-pretrained models such as support vector machines and boosting.
\end{abstract}

\textbf{Statement of Significance}

Answers to open-ended questions are now increasingly coded automatically using statistical learning models. Computer scientists have found that pre-training language models on vast amounts of unrelated data (e.g., Wikipedia) and then fine-tuning them to the text data of interest can improve the accuracy of classifying text into categories. However, language models are typically fine-tuned on millions of words, whereas answers to open-ended questions are typically much more limited. We demonstrate that the pre-trained model ``BERT'' nonetheless improves accuracy for coding answer to open ended questions.

\section{Introduction}

Open-ended questions are important because, unlike closed-ended questions, open-ended questions allow respondents to express their opinions without constraints and thus may better capture respondents' thoughts \citep{bengston11,he21}. Answers to open-ended questions are text data.
They are often coded manually into different classes by human experts. This is time consuming, inconvenient \citep{geer91}, and does not scale well in large survey data.

An alternative is automatic coding. Here, statistical models are trained on a small subset of manually coded data which is then used to predict the remaining unlabeled answers. There are typically two tasks: (a) converting text answers into numerical variables to be used in statistical modeling and (b) fitting a classification model to the converted data. For task (a), original text answers are converted into numerical variables from which classification algorithms can learn their relationships with codes or classes.
A popular conversion approach is ngram representation \citep{Schonlau17} in which each input text is converted into a sparse vector. Each element in the vector indicates the presence or the frequency of a distinct word in the input text. The resulting vector is high-dimensional (number of distinct words in all texts) and sparse (most elements are zero) because most words do not appear in any one input text.
Limitations of ngram representation in text classification include lack of contextual information and difficulty in handling words unseen in the training data. Another approach is word embedding in which each word is represented as a dense numerical vector in a lower dimensional space.

For task (b), the numerical variables are fed to a statistical learning model such as Support Vector Machines \citep{Vapnik2000} or Random Forests \citep{breiman01} for model training. The model is trained on roughly a couple of hundred manually coded answer texts. The trained model can then be used to predict (i.e., automatically code) the remaining text answers.

The use of statistical learning methods for analyzing text data from open-ended questions has received increasing attention in social sciences \citep{hill19,conrad21}. Because data quality plays a central role in the social sciences, \citet{schonlau2016semi} proposed using a method called semi-automatic classification to code open-ended responses, where any answers with a predicted coding accuracy below a threshold probability are coded manually.

Answers that can be coded into more than one category, similar to check-all-that apply questions in surveys and referred to as multi-label answers, are hard to code because the number of labels is not known, and each of the labels has to be classified correctly.
\citet{schonlau2021all-that-apply} investigated automatic classification of multi-label answers using standard algorithms; \citet{gweon20} developed a semi-automatic classification algorithm for multi-label answers.

Occupation coding refers to classifying a respondent’s occupation based on their answer to an open-ended question (``What is your main job?''). Occupation coding is a special case because it involves a large number of possible classes (e.g., 700) and the classes are structured hierarchically with job titles nested based on their relevance or similarity. Also, the respondent's answer consists usually only of a few words rather than a sentence or two.
Efforts for automatic occupation coding include original efforts to exploit  that hierarchical nature \citep{gweon17},
and a comparison of many different statistical learning algorithms and original modifications \citep{schierholz21}.

In addition to automated text classification, statistical learning methods have also been widely used for other survey problems such as predicting nonresponse in panel studies \citep{Kern2021} and adjusting nonprobability samples for improving their population representativeness \citep{kern2020}.

In recent years, the community of natural language processing (NLP) has come to realize that is useful to pre-train a so-called language model that ``understands'' language, i.e. that can predict missing or following words. A language model is first trained on a large corpus of texts, and then the model is adapted to a specific classification or other tasks for the specific domain of interest. This is an example of transfer learning, because we transfer the knowledge about general language learned elsewhere to a specific domain of interest. For language models, this form of transfer learning is called pre-training.

The currently most popular pre-trained language model is BERT (Bidirectional Encoder Representations from Transformers) \citep{devlin18}. BERT is pre-trained on big data sources from BookCorpus and Wikipedia which have in total more than 3.5 billion words. Recent studies have shown that BERT achieves the state of the art performance in many NLP tasks including text classification \citep{kaliyar2021,cunha21,gasparetto22}.

Given that BERT's parameters are estimated during pre-training, it is possible to use BERT in two ways: with frozen parameters, and with fine-tuned parameters.
Frozen parameters mean that the pre-trained parameters do not change during domain-specific training. In this situation, BERT relies on the general context of a language. Fine tuning takes the frozen parameters as starting values, but allows them to change while training on the domain-specific data set. In both cases, additional parameters needed to convert BERT to the specific task of interest have to be estimated from scratch.

Many applied case studies demonstrate the effectiveness of fine-tuning when the domain-specific data set is large.
For example, the developers of BERT \citep{devlin18} tested fine-tuning BERT for text classification using a collection of text data, the General Language Understanding Evaluation (GLUE) benchmark \citep{wang18}, that consists of data sets of size ranging from 2,500 to 392,000 where a unit of observation ranges from a news article to a book.
Also, \citet{sun19} focused particularly on the text classification of fine-tuned BERT on 8 benchmark text data sets of sizes ranging from 5,000 to 1,400,000 observations.
Such data sizes are large when compared to the number of survey respondents that answered an  open-ended question. It is not clear whether fine-tuning on smaller domain specific data sets would not work as well, e.g., due to numerical instabilities.

While results with BERT are clearly impressive, it is not clear to what extent such results carry over for classifying open-ended questions in surveys.
One, we are interested in training on a few hundred examples at the most. These are much smaller data sets than BERT is commonly used with.
Two, the classification for open-ended questions is a different domain of application which may not lend itself as easily to BERT model: perhaps open-ended answers are not as easily classifiable due to greater ambiguity, or because they require a greater level of abstraction than the oft-used Reuters news data.

In this paper, we investigate the effectiveness of BERT for the automated coding of open-ended questions. We compare BERT to the traditional alternative, training on ngram variables, for very small training data sets.

\section{\label{s:bert}The BERT model for text classification}

This section gives a high-level view of the BERT model and the developments leading up to it: dense word embeddings and the Transformer.

\subsection{Word embedding and Transformer}

About 10 years ago, the natural language processing community developed  word embedding techniques that efficiently represent words
by dense numerical vectors \citep{khattak19,wang20}.
 The relationships between vectors of words mirror the linguistic relationships of the words. Words with different meanings (e.g., words `good' and `bad') lie far away from each other in the embedded space, whereas words with similar meanings (e.g., words `good' and `great') lie close to each other.
Popular word embedding algorithms include Word2Vec \citep{mikolov13} and GloVe \citep{pennington14}.

Despite their ability to capture semantic relationships between words, the early word embedding methods are static: Each word has a fixed position regardless of its contextual meaning in a sentence \citep{wang20-2}.
Static embeddings cannot accommodate words with multiple meanings (e.g. lead, row, bow) where the meaning depends on the context in a sentence.
The introduction of the Transformer \citep{vaswani17} allows for context dependent embeddings, that is, how a word is converted (or embedded) to a  vector changes depending on the surrounding words.

The Transformer is intended for translation from one language to another and consists of two parts: the encoder embeds words from one language into dense vectors. The dense vectors no longer depend on the language, which enables the decoder to decode the vectors into words of a different language. Importantly, the meaning of words is encoded as vectors.

\subsection{The BERT model}

While originally intended for translation, the Transformer architecture \citep{vaswani17} has been adapted to  build language models. A language model is a model that can predict subsequent words in  a sentence or a paragraph.
BERT (Bidirectional Encoder Representations from Transformers) \citep{devlin18} is a language model that uses the encoder layers of the Transformer.
To train an encoder for prediction,
BERT  incorporated cloze tests to train the model, i.e. it removed a portion of words that then had to be predicted. The cloze tests were given a different name: masked language modeling.
Masked language modeling considers both the right and left contexts of a masked word jointly and it was argued that this bi-directional approach is superior to a single-direction approach, where the next word is predicted without knowledge of which words follow the next word.
In addition to masking, BERT also uses a secondary task, next sentence prediction, where BERT is given two sentences and predicts whether or not the two sentences follow each other.

BERT was introduced with two versions. BERT-base has  110 millions of parameters. BERT-large has 340 millions of parameters.
Both models were pre-trained on BookCorpus and Wikipedia which have in total more than 3,500M words.

The pre-trained BERT model is considered to have a general sense of a language (e.g., being aware of context and word ordering in a given text).
For text classification, an additional classification layer is added on top of BERT's output layer, and the model parameters are fine-tuned based on a set of given data of interest.
The fine-tuning stage includes learning the new classification layer parameters and modifying all the pre-trained parameters of BERT.

\section{Simulation study} \label{experiments}

\subsection{Data set description}

{\bf Patient Joe.} The Patient Joe data set was collected to study to what extent patients
are actively engaged in their own health care.
Participants for the New England Family Study (NEFS) were recruited between 2004 and 2007 as  described in more detail in \citep{martin11}. 800 individuals were selected from NEFS through a multi-stage sampling plan, of which 614 (RR2=77\%) were interviewed. Of these, 585 individuals answered the probe (cumulative 73\%).

As part of a larger study, respondents were asked to answer the following open-ended question \citep{martin11,Schonlau17}: ``Joe’s doctor told him that he would need to return in two weeks to find out whether or not his condition had improved. But when Joe asked the receptionist for an appointment, he was told that it would be over a month before the next available appointment. What should Joe do?''  Answers were recorded on audio-tape and subsequently transcribed. Answers were labeled by two raters (Kappa= 80\%). Answers were classified into exactly one of the four categories: proactive (51.5\%),
somewhat proactive (10.1\%),
passive (33.3\%) and
counterproductive (5.1\%).
The data set contains 585 observations. The text answers have a median length of 22 words (25th percentile: 13 words, 75th percentile: 35 words).
The data are available from \citet{patientjoe22}.

{\bf Disclosure.}
The disclosure data set was collected to learn the relationship between risk of disclosure and survey participation.
The sample was drawn from Survey Sampling International's Internet panel. The participation rate was 2.1\%.  The original paper \citep{Couper08} does not mention the survey dates.

A series of vignettes describing different levels of disclosure risk was randomly assigned to respondents \citep{Couper08}. Respondents were then asked how likely they would be to participate in the described survey.
Respondents were then asked an open-ended probe: ``Why would you participate?". Our analysis focuses on this question.
Answers were labeled by two raters (Kappa=79\%).
The disclosure data
\citep{disclosure16}
are available as part of the replication materials for \citet{schonlau2016semi}.

A total of 1,212 positively worded answers were classified into exactly one of the eleven categories
as shown in Table~\ref{tab1}. The text answers have a median length of 10 words (25th percentile: 5 words, 75th percentile: 16 words).

\begin{table*}[htp]
\centering
\caption{Classes in the disclosure data. Rare classes were combined and labeled as ``Others''.}
\label{tab1}
\begin{tabular}{lrl} \toprule
Class & Data fraction & Description \\ \midrule
C1 & 0.023  & Believes in research generally \\
C2 & 0.278 &Wants to be helpful / express opinion \\
C3 & 0.144 & I like surveys, I enjoy surveys \\
C4 & 0.034  & I’d learn something, interested in results, curiosity \\
C5 & 0.140 & The money (incentive) \\
C6 & 0.134 & Topic-related (interesting, issues important) \\
C7 & 0.045 & Survey doesn’t take much time, short survey \\
C8 & 0.020  & Other survey characteristic \\
C9 & 0.054 & No objection \\
C10 & 0.043 & Uncodable response, ambiguous \\
C11 & 0.086 & Others \\
\bottomrule
\end{tabular}
\end{table*}

\subsection{Results}

We compared BERT against other popular machine learning algorithms: random forest (RF) \citep{breiman01}, XGBoost (XGB) \citep{chen16} and support vector machines (SVM) \citep{Vapnik2000}. All models used the same (random) sets of training/validation/test data. Hyperparameter tuning details are found in Appendix A.

Figure~\ref{result1} presents the classification accuracy of each method on the Patient Joe (4 classes) and Disclosure data (11 classes).
The classification accuracy is the percentage of texts for which the classes were correctly predicted (in the test data).

When the training data contained 100 observations, we observed small differences in classification performance between BERT and the other classifiers.
As the size of training data increased, the fine-tuned BERT outperformed the others and the pattern was observed in both data sets.
For the Patient Joe data with 400 training examples, BERT achieved classification accuracy of 0.854 and the second best result achieved by XGBoost was 0.785.
For the Disclosure data, the fine-tuned BERT achieved accuracy of 0.725 and the other algorithms performed almost the same with an accuracy of 0.618.
The ngram-based models showed similar performance on both data sets.
The BERT model without fine-tuning performed the worst. With 400 training observations, the classification accuracy difference between with/without fine-tuning was over 20 percentage points on both data sets, a large difference.

\begin{figure}[H]
\begin{subfigure}[t]{.47\textwidth}
\centering
\includegraphics[scale=0.38]{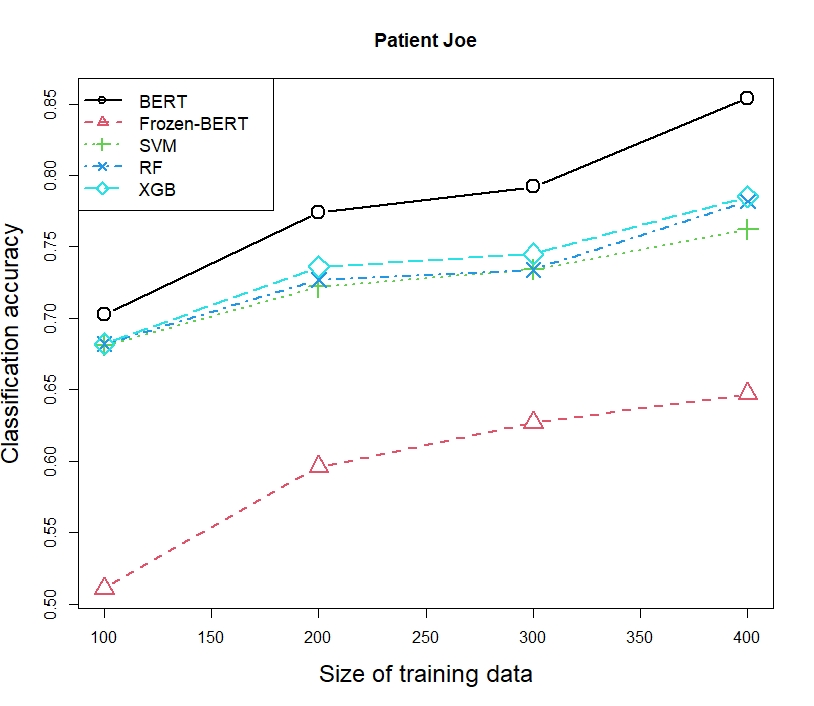}
\end{subfigure}
\begin{subfigure}[t]{0.47\textwidth}
\centering
\includegraphics[scale=0.38]{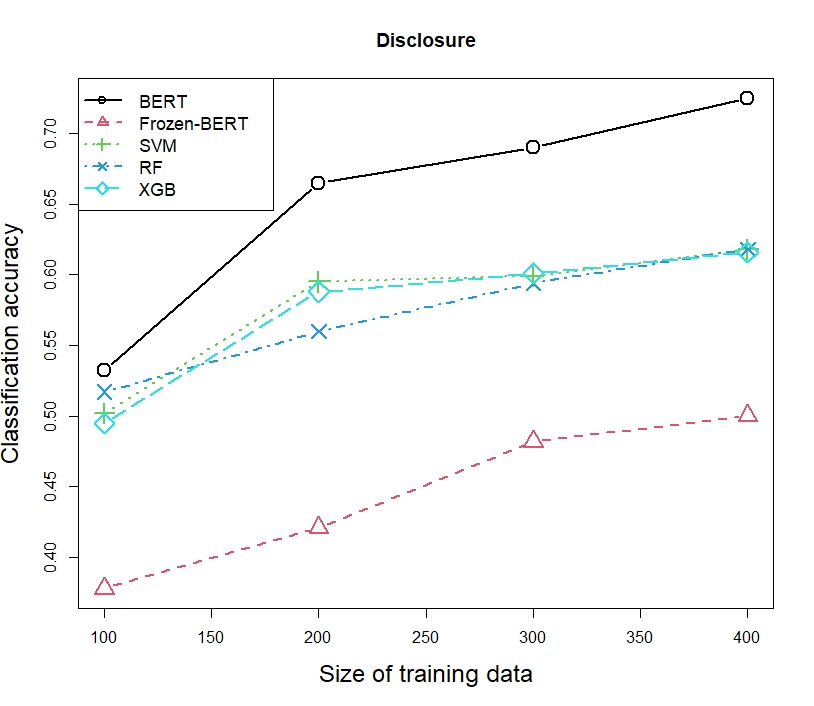}
\end{subfigure}
\caption{Classification accuracy of each method for the Patient Joe (left) and Disclosure (right) data}
\label{result1}
\end{figure}

\begin{figure}[H]
\begin{subfigure}[t]{.47\textwidth}
\centering
\includegraphics[scale=0.38]{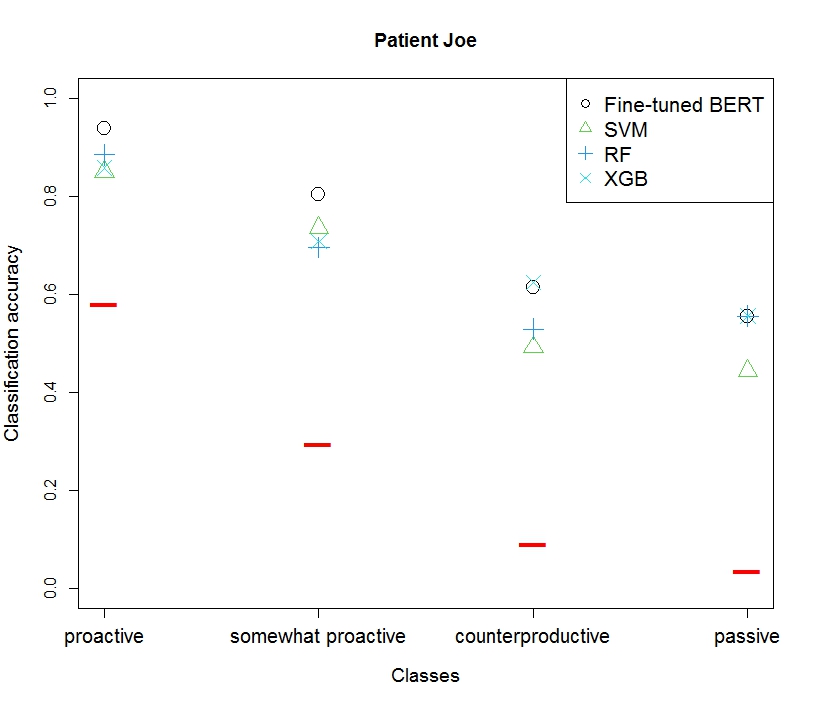}
\end{subfigure}
\begin{subfigure}[t]{0.47\textwidth}
\centering
\includegraphics[scale=0.38]{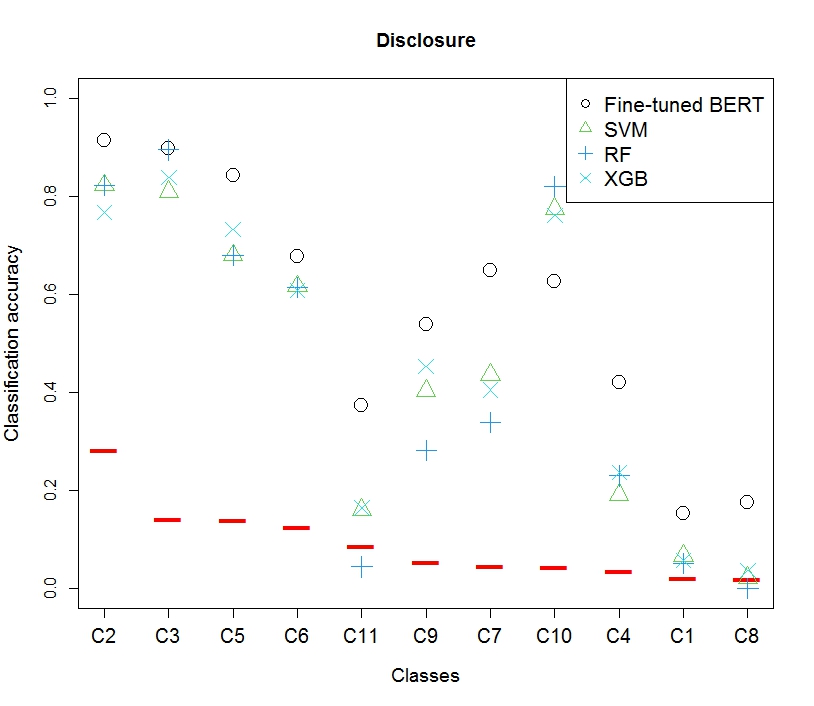}
\end{subfigure}
\caption{Class-wise accuracy of each method for the Patient Joe (left) and Disclosure (right) data with 400 training examples. Each horizontal solid bar represents the relative frequency of the corresponding class. The classes are ordered by their frequencies.}
\label{result2}
\end{figure}

The results in Figure~\ref{result2} show the accuracy by class at training data of size 400.
For the Patient Joe data, the fine-tuned BERT was more accurate than the others mostly on the two dominant classes (proactive and somewhat proactive).
For the Disclosure data, the fine-tuned BERT outperformed the others for 10 out of 11 classes, regardless of their frequencies.

\section{Discussion\label{s:discussion}}

With 100 training observations, the ngram-based models (SVM, RF and XGB) and the fine-tuned BERT showed small differences in classification performance for both Patient Joe and Disclosure data sets. Relative to the ngram based models, the fine-tuned BERT became more effective as the size of training data increased.

Fine-tuning vastly improved the classification accuracy of BERT (by 10 -- 20 percentage points). BERT without fine-tuning was beaten by the ngram-based models. This suggests that the pre-trained language model requires fine-tuning even for very small training data sets.

Both data sets from open-ended questions have imbalanced class distributions.
The class-wise accuracy results show BERT's better classification performance relative to ngram-based models is not particularly biased towards certain major classes.
BERT tends to achieve higher accuracy for most classes without a noticeable drop in performance for any one class. We also examined the methods in terms of  macro-accuracy (i.e., the average of the class-wise accuracy) that values all categories equally regardless of their frequencies. The results were similar to those we had obtained from overall accuracy.

We see the following limitations:
First, for 400 training observations, the accuracy in the two examples is roughly 75-85\%. While considerably improved, this may not be enough for some purposes. If so, we recommend semi-automatic classification \citep{schonlau2016semi,gweon20}, where there is a threshold probability below which answers are coded manually.
Also, BERT is harder to program than non-pre-trained methods. At present, most analysts are programming in Python to access the pre-trained BERT model. It is also computationally intensive as fine-tuning requires at least a high-end laptop or online resources such as Google Colab.
Lastly, BERT has been very successful in English. BERT models in other languages are being developed. While we are hopeful these results will extend to other languages, this is not certain.

The social sciences emphasize data quality.
While clearly useful,  models such as BERT still require  close scrutiny.
Nonetheless, we see some advantages of fine tuning a BERT model even for classifying just a few thousand observations. One, automatic classification makes classification results reproducible (though they still depend on the initial training data). Two, in panel studies there may be left over data from a previous wave that can be used as training data. Three, survey data sets may be getting larger and we want to build analysis  capacity. Four, whereas 500 training observations may be classified in house, classifying 3000 observations may require hiring coders to do this work manually.
We conclude that the use of BERT is a promising tool for automatic classification of open-ended questions as compared with non-pre-trained classification methods.
The advantage of BERT relative to non-pre-trained methods increases as the more data are used for fine-tuning.

\section*{Acknowledgement}
This study design and analysis was not preregistered.

\section*{Appendix A: Models and hyperparameter tuning}

For each data set, each model was trained on  training data of size from 100 to 400. For the non-BERT models, the model parameters were tuned using validation data of size 100. The remaining observations were used for testing the performance of each model.

We used the uncased BERT base model (12 Layers, 768 hidden nodes, 12 attention heads).
We specified cross-entropy loss and used 10\% dropout.
For fine-tuning the BERT model, we used epochs = 3, batch size = 8, and learning rate = 0.0003. The dropout probability was set to 0.1.
For the non-BERT approaches, text answers were converted into unigram variables for model training.
The parameters of SVM were tuned with a grid search for $C \in \{0.1, 1, 10\}$, $\gamma \in \{0.01, 0.1, 1\}$ and kernel $\in \{`rbf', `linear', `poly'\}$.
For tuning the parameters of RF, we used a grid search for mtry $\in \{0.3, 0.5, 1\}$, minimum number of samples at each terminal node $\in \{1, 3, 5\}$ and number of trees $\in \{300, 600, 900\}$.
The XGB model has more tuning parameters than SVM and RF. For efficient parameter tuning, we used a random search for learning rate $\in \{0.01, 0.1, 0.3\}$, maximum depth $\in \{3, 5, 10\}$, number of estimators $\in \{100, 200, 300\}$, $\gamma \in \{0, 0.1, 1\}$, $\lambda \in \{0.3, 0.5, 0.7\}$, $\alpha \in \{0, 0.1, 0.2\}$ and sub-sample proportion $\in \{0.5, 0.7, 1\}$. We chose the best combination from 50 random iterations. The python code is available in the online appendix.

\bibliography{reference_bert}
\bibliographystyle{chicago}

\end{document}